\documentclass[]{aa}

\usepackage{graphicx}
\usepackage{amssymb}
\usepackage{latexsym}
\usepackage{amsmath}
\usepackage{txfonts}

\newcommand{\ds}{\displaystyle}
\newcommand{\du}{\mathrm{d}}
\newcommand{\expo}{\mathrm{e}}
\newcommand{\mtot}{\mathrm{E}}
\newcommand{\meps}{\mathrm{E}_{\epsilon}}

\begin{document}
\title{Single-Mode versus multimode interferometry: \\ a performance study}
\author{E. Tatulli, P. M\`ege, A. Chelli}
\institute{Laboratoire d'Astrophysique, Observatoire de Grenoble, 38041 Grenoble cedex France}
\titlerunning{Single-mode interferometry}
\offprints{Eric Tatulli, \email{Eric.Tatulli@obs.ujf-grenoble.fr}}
\date{Received /
  Accepted}
\abstract{We compare performances of ground-based
  single-mode and multimode (speckle) interferometers in the presence of partial Adaptive
  Optics correction of atmospheric turbulence. It is first
  shown that for compact sources (i.e. sources smaller than the
  Airy disk of a single telescope) not entirely resolved by the interferometer, 
  the remarkable property of spatial filtering of single-mode 
  waveguides coupled with AO correction significantly reduces 
  the speckle noise which arises from residual wavefront
  corrugations.
  Focusing on those sources, and in
  the light of the AMBER experiment (the near infrared instrument of
  the VLTI), we show that single-mode interferometry produces 
  a better Signal-to-Noise Ratio on the visibility than speckle
  interferometry. This is true for bright sources ($K < 5$), 
  and in any case    
  as soon as Strehl ratio of $0.2$ is achieved.
  Finally, the fiber estimator  
  is much more robust -- by two orders
  of magnitude -- than the speckle estimator 
  with respect to  Strehl ratio variations during the
  calibration procedure. The present analysis theoretically explains why
  interferometry with fibers can produce visibility
  measurements with a very high precision, $1\%$ or less.
}
\maketitle


\section{Introduction}
The great interest of using spatial filtering properties of optical
waveguides in astronomical interferometers has been proven in the past
years (\cite{coude_du_foresto_etal_1}, \cite{berger_1}). As a
consequence, integrated optics and fibers are more and more introduced
in the design of present and future interferometers to carry the
signal to the detector. Furthermore, practical and theoretical studies
(\cite{haguenauer_1}, \cite{guyon_1},  \cite{mege_3}) have been undertaken to
investigate the physical and optical properties in waveguided
interferometers.

The present work aims at comparing the sensitivity and robustness
of single-mode and multimode (speckle) interferometry. In Section
\ref{sec_modvis}, we recall the basic concepts of the fibered interferometric
equation and the modal visibility. We derive, in Section
\ref{sec_error_vis}, the formal expression of the Signal to Noise Ratio
(SNR) of the modal visibility which takes into account photon, detector and
atmospheric noise. In Section \ref{sec_pseudo_speck}, we
propose an analytical approach to estimate the profile of the
visibility SNR as a function of the magnitude, from partially Adaptive
Optics (AO) corrected interferograms. We also derive the
performance of single-mode interferometry applied to the AMBER experiment
(the near infrared instrument of the VLTI), in the case of single Gaussian sources. Finally, in Section \ref{sec_multi},
we compute the performances of the multispeckle method (\cite{roddier_lena_1},
\cite{mourard_etal_1}) currently used to 
estimate visibility from non-fibered interferometers and we
compare the performances and the robustness of 
single-mode and speckle interferometry.

\section{The modal visibility} \label{sec_modvis}
M\`ege (\cite{mege_1}) theoretically described  how the light is carried
and processed through an interferometer with optical waveguides. Specifically, he highlighted the coupling phenomenon between the incoming
wavefront and the fiber, and analyzed the characteristics of the
interferogram recorded on the detector. The principal results concerning the 
interferometric equation of a  N telescopes ($N_{tel}$) fibered interferometer and the resulting modal
visibility can be summarized as following:
\begin{eqnarray}
I(f)&=&\sum_i^{N_{tel}} K_i \rho_i(V_{\star}) H_i(f) +\nonumber \\
&&\sum_i^{N_{tel}}\sum_j^{N_{tel}} \sqrt{K_iK_j}  \rho_{ij}(V_{\star})  H_{ij}(f-f_{ij}) \label{eq_interf}
\end{eqnarray} 
where $I(f)$ is the Fourier transform of the interferogram at the spatial
frequency $f$, $V_{\star}$ is the visibility of the source and $K_i$ is
the number of photoevents from the $i^{th}$ telescope that would be detected 
in absence of fibers. $H_i(f)$ and $H_{ij}(f-f_{ij})$ are the Fourier
transforms of the so-called normalized
carrying waves centered at respectively the frequencies $f=0$ and $f=f_{ij}$
(\cite{mege_4}). Their shape, hence the shape of the interferogram, is entirely
determined by the geometry of the fibers.
$\rho_{i}(V_{\star})$ and $\rho_{ij}(V_{\star})$ are respectively the
low- (LF) and high-frequency (HF)
instantaneous coupling coefficients. They give the fraction of flux entering in the
fibers respectively for the photometric and the interferometric
channels. Their theoretical expressions are:
\begin{eqnarray}
\rho_i(V_{\star})&=&\rho_0 \frac{(V_{\star}*T^i)_{f=0}}{\int T^{i}_0(f) \du{f}}\label{eq_rhoi}  \\ 
\rho_{ij}(V_{\star})&=&\rho_0 \frac{(V_{\star}*T^{ij})_{f=f_{ij}}}{\sqrt{\int T^{i}_0(f) \du{f}\int T^{j}_0(f) \du{f}}} \label{eq_rhoij}
\end{eqnarray}
where $T^{i}$ and $T^{ij}$ are the (partially AO corrected) modal
transfer functions resulting respectively from the normalized
auto-correlation and cross-correlation of the entrance
aberration-corrupted pupil weighted by the fiber single mode
(\cite{roddier_1}, \cite{mege_4}). $T^{i}_0$ is the turbulence-free
modal transfer function and $\rho_0$ is the optimum coupling
efficiency fixed by the fiber core design
(\cite{shaklan_1}). Equations \ref{eq_rhoi} and \ref{eq_rhoij}
generalize the turbulent coupling efficiency to any kind of
sources, i.e. not only when it is unresolved by a single
telescope (as first noticed by
\cite{dyer_1} in the non-turbulent case). Note that, for a point source, the
low frequency coupling coefficient is proportional to the Strehl ratio
 $\mathcal{S}$, $\rho=\rho_0\mathcal{S}$ (\cite{foresto_1}), 
and the high-frequency coupling coefficient follows the simple
relationship $|\rho_{ij}|^2=\rho_{i}\rho_{j}$.  

From Eq.'s 
\ref{eq_rhoi} and \ref{eq_rhoij}, we can deduce the expression of the
instantaneous modal visibility $V_{ij}^2$ at the spatial frequency
$f_{ij}$. It is defined as the 
ratio between the coherent energy (high frequency) and the incoherent one (low frequency):
\begin{equation}
V_{ij}^2 =
\frac{|\rho_{ij}(V_{\star})|^2}{\rho_i(V_{\star})\rho_j(V_{\star})} \label{eq_vis_definition}
\end{equation}
Perfect equality between the instantaneous modal visibility and the
  object visibility exists only for 
  point sources ($V_{ij}^2 = V_{\ast}^2 = 1$). In the general case,
  however, the instantaneous modal visibility does not match that of
  the object, especially if the source is extended. 

In terms of measurements, the estimator of the modal visibility
$\widetilde{V_{ij}^2}$ can be computed as the ratio between 
the interferogram power spectrum at the frequency $f_{ij}$:
$|I^2(f_{ij})|= (1-\tau)^2|\rho_{ij}(V_{\star})|^2K_iKj$, and the
photometric fluxes: $k_i=\tau\rho_i(V_{\star})K_i$, where $\tau$
is the fraction of light selected for photometry at the
output of the beam-splitter. Assuming that the
latter are estimated independently through dedicated outputs of the optical
waveguide (so-called photometric outputs), it holds:
\begin{equation}
\widetilde{V_{ij}^2}
=\frac{<|I^2(f_{ij})|>}{<k_ik_j>}\left(\frac{\tau}{1-\tau}\right )^2 \label{eq_vis_estimate}
\end{equation}

In the next Section, we derive a formal expression of the relative error --
the inverse of the SNR -- of the modal visibility. 

\section{Relative error of the modal visibility}\label{sec_error_vis}
The noise calculations are based on the spatially continuous model of
photodetection introduced by Goodman \cite{goodman_1}, (see also \cite{chelli_1},
\cite{berio_etal_2}). Within this framework, the
signal is assumed to be corrupted by three different types of noise: (i) the photon
noise, with $(1-\tau)K$ and $\tau K$ the total number of detected
photoevents in the interferometric and the photometric channels respectively; (ii) the
additive Gaussian noise of global variance $\sigma^2$
which arises from the detector and from thermal emission; (iii) the atmospheric
noise which results from the coupling efficiency variations due to the
turbulence.

The noise contributions are derived in Appendix \ref{app_noise} and
\ref{app_vis_estimator}. The square of the relative error of the modal 
visibility can be described as the sum of three terms:
\begin{equation}
\frac{\sigma^2\{V_{ij}^2\}}{\overline{V_{ij}^2}^2} = \mathcal{E}_P^2(K,\rho) 
+ \mathcal{E}_A^2(K, \sigma^2, \rho) + \mathcal{E}^2_S(\rho) \label{eq_modvis err},
\end{equation}
where $\mathcal{E}_P(K,\rho)$, $\mathcal{E}_A(K,\sigma^2,
  \rho)$ and  $\mathcal{E}_S(\rho)$, are the relative errors due to
  photon, additive and atmospheric noise, respectively. It holds:
\begin{eqnarray}
\mathcal{E}_{P}^2 &=&\left[\frac{N_{tel}(4\overline{\rho_{lf}|\rho_{ij}|^2}-2\overline{\rho}\overline{|\rho_{ij}|^2})}{(1-\tau)\overline{|\rho_{ij}|^2}^2}
  +\frac{2}{\tau}\frac{\overline{\rho^2}}{\overline{\rho}^3}\right]\frac{N_{tel}}{\overline{K}}
\nonumber \\
& + &
\left[\frac{N_{tel}^2(2\overline{\rho_{lf}^2}-\overline{\rho}^2)}{(1-\tau)^2\overline{|\rho_{ij}|^2}^2}
  +\frac{4}{(1-\tau)^2\overline{|\rho_{ij}|^2}}
  +\frac{1}{\tau^2\overline{\rho}^2}\right]
\frac{N_{tel}^2}{{\overline{K}^2}}\nonumber \\
& + & \frac{\overline{\rho}}{(1-\tau)^3\overline{|\rho_{ij}|^2}^2} \frac{N_{tel}^4}{{\overline{K}^3}}
\label{eq_snr_P}\\
\mathcal{E}_{A}^2 &=& \frac{3N_{pix}\sigma^4 + N_{pix}^2\sigma^4}{(1-\tau)^4\overline{|\rho_{ij}|^2}^2}\frac{N_{tel}^4}{\overline{K}^4}
+
\frac{2N_{pix}\sigma^2}{(1-\tau)^2\overline{|\rho_{ij}|^2}}\frac{N_{tel}^2}{\overline{K}^2}
\nonumber \\
&+& \frac{2N\sigma^2\overline{\rho}}{(1-\tau)^3\overline{|\rho_{ij}|^2}^2}\frac{N_{tel}^4}{\overline{K}^3}\label{eq_snr_A}\\
\mathcal{E}_{S}^2 &=&\frac{\sigma^2_{|\rho_{ij}|^2}}{\overline{|\rho_{ij}|^2}^2}+\frac{\sigma^2_{\rho_i\rho_j}}{\overline{\rho}_{i}\overline{\rho}_{j}}-2\frac{\mathrm{Cov}\{|\rho_{ij}|^2,\rho_i\rho_j\} }{\overline{|\rho_{ij}|^2}\overline{\rho}_{i}\overline{\rho}_{j}} \label{eq_specknoise}
\end{eqnarray}
where $\overline{X}$ denotes the expected value of the random quantity
  $X$, and $\mathrm{Cov}\{X,Y\}$ the covariance between $X$ and
  $Y$. $\rho_{lf}$ is the average over all the telescopes of the $LF$ coupling
coefficient, and $N_{pix}$ is the number of pixels per
  interferogram (see Appendix \ref{app_vis_estimator} for more details). 

For bright sources, with the exception of point sources for which
$|\rho_{ij}|^2=\rho_{i}\rho_{j}$ (and hence $\mathcal{E}_S(\rho) =0$), 
the dominant noise is the atmospheric noise, which results from the
classical speckle noise filtered by the fiber.  From now on, we refer to
that noise as the modal speckle noise, and the corresponding SNR, 
($\mathcal{E}_S(\rho)^{-1}$), as the modal speckle SNR. This depends on
the variation of the $LF$ and
$HF$ coupling coefficients and hence, on the strength of the
turbulence. As in the speckle case, it does not depend on the source brightness and
corresponds to the maximum achievable SNR per interferogram. Nevertheless, as we will show,
it depends on the source size. 
\begin{figure*}[!t]
\begin{center}
\includegraphics[width=17cm,height=11cm]{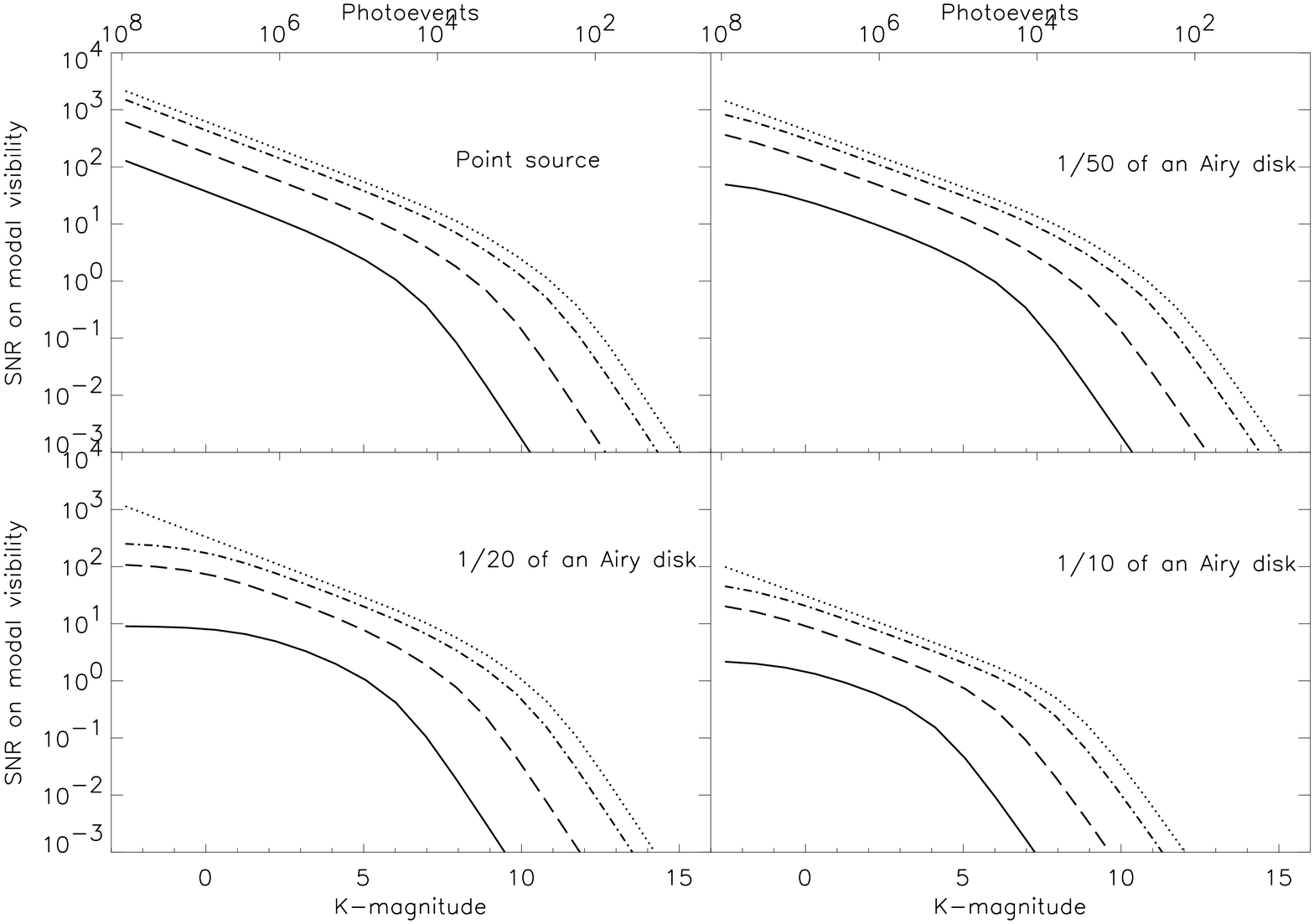}
\caption{\label{fig_modal_vis_snr}\footnotesize{Modal SNR per
    interferogram as a function
    of the magnitude of the source in the $K$ band. From left to
    right, top to bottom, the object size increases:
    point source, $1/50$, $1/20$, $1/10$ of Airy disk. The considered
    baseline is 
    $Bx=By=100\mathrm{m}$ which mimics an average baseline of the VLTI
    array. This corresponds 
    to visibility of respectively $1$, $0.882$, $0.457$, $0.04$.  
Results are given for different Strehl ratios:
    $\mathcal{S}=0.011$ (solid line, no correction),
    $\mathcal{S}=0.1$ (dashed line), $\mathcal{S}=0.5$ (dash-dotted
    line) and $\mathcal{S}=0.9$ (dotted line).}}
\end{center}
\label{modal_snr}
\end{figure*}
\section{Performances of single-mode interferometry}
\label{sec_pseudo_speck}
In this Section, we  develop a simple model to estimate the
modal speckle noise from partially AO corrected interferograms, and we derive the
performances of single-mode interferometry applied to AMBER, the near-infrared instrument of the Very Large Telescope Interferometer (VLTI). 
\subsection{The modal speckle noise}
The derivation of the modal speckle noise is fully detailed in Appendix
  \ref{app_pseudo_speck}: here we recall the outlines of our approach. We assume that: (i) the distance between
  the telescopes (baseline) is larger than the
  outer scale of the turbulence, i.e. that the complex
  amplitudes over pupils $i$ and $j$ are uncorrelated; (ii) the
  atmospheric phase has Gaussian statistics; (iii) the associated structure
  function is (spatially) stationary.  

 
At this stage, the modal speckle SNR is described by integrals of
  dimensions up to sixteen (see  Table
\ref{table_exp_coupling}, from Eq \ref{eq_app_fib1} to Eq 
\ref{eq_app_fib5}), which depend on the pupil, the object visibility
and the structure functions, respectively. Here we estimate these integrals with a simplified analytical approach dealing
with Gaussian functions only.
We assume that the object
brightness distribution $O_{\ast}(\alpha)$ and the pupil function 
have respectively the form:
\begin{equation}
O_{\ast}(\alpha) = \exp\left(-{\alpha^2 \over \sigma_o^2}\right) \label{eq_obj}
\end{equation}
and,
\begin{equation}
P(f) = \exp\left(-\frac{f^2}{\sigma_P^2}\right),
~~\sigma_P=\frac{1}{\sqrt{2}}\frac{D}{\lambda} \label{eq_pupfunc}
\end{equation}
where $D$ is the diameter of the telescope and $\sigma_p$ is defined 
in such a way that the integral of the Gaussian pupil, i.e. its surface, is equal to the integral of a circular pupil of diameter $D$.
Furthermore, 
in order to take into account the partial correction by adaptive optics, we
approximate the structure function $\mathcal{D}(f)$ by:
\begin{equation}
\exp\left[-\frac{1}{2}\mathcal{D}(f)\right] = h+(1-h)\mathcal{B}_{\Phi}(f) \label{eq_struct_func}
\end{equation}
where $h \in [0,1]$ defines the level of correction and 
$\mathcal{B}_{\Phi}(f)$ is the transfer function of the turbulent
atmosphere (\cite{conan_1}) that we will assume Gaussian too:
\begin{equation}
\mathcal{B}_{\Phi} =
\exp\left(-\frac{f^2}{\sigma_{\mathcal{B}}^2}\right),
~~\sigma_{\mathcal{B}}=\sqrt{\frac{2}{6.88}}\frac{r_0}{\lambda} \label{eq_bfunc}
\end{equation}
$r_0$ being the Fried parameter. Note that, from the previous equations, the Strehl ratio $\mathcal{S}$ is given by:
\begin{equation}
\mathcal{S} = h + (1-h)\frac{\sigma_{\mathcal{B}}^2}{2\sigma_P^2+\sigma_{\mathcal{B}}^2}
\label{eq_strehl}
\end{equation} 
At this point, the integrals \ref{eq_app_fib3}, \ref{eq_app_fib4} and 
\ref{eq_app_fib5} cannot be computed formally yet. To do that, we perform a
limited expansion of the expressions under the integrals with respect to
the variable $\epsilon = (1-h)/h$. We end up with a series development
in $\epsilon^n$ in which the coefficients are integrals of Gaussian functions
products only. These coefficients are then computed with the MAPLE
software. However, the series development converges only for $\epsilon \in [0,1[$
  or $h \in ]0.5,1]$, i.e. for good to perfect AO corrections (see Eq. 
\ref{eq_strehl}). In order to estimate the modal speckle SNR for average
  to low AO correction, we compute it in the pure turbulent case ($h=0$) and we
  extrapolate from $h=0$ to $h=0.5$. The case $h=0$ is computed
  separately from standard hypothesis assuming that the complex amplitude of the pure 
  turbulent wavefront follows circular Gaussian statistics (\cite{roddier_1}).   
We finally obtain an expression of the modal speckle SNR as a function of
  the major parameters of the observation: the source
size, the baseline, the turbulence strength $D/r_0$,
 and the level of AO correction, i.e. the Strehl ratio $\mathcal{S}$. 
Strictly speaking, the structure function is not stationary
as the error of the AO corrected wavefront increases
from the center to the edge of the telescope pupil. Also, its
shape is not exactly described by the simplified
Eq. \ref{eq_struct_func}. Nevertheless, we expect the modal speckle SNR 
resulting from our model, dealing with Gaussian functions only, to have
the right order of magnitude and the correct functional dependencies. 
\begin{figure}[!t]
\begin{center}
\includegraphics[width=\columnwidth]{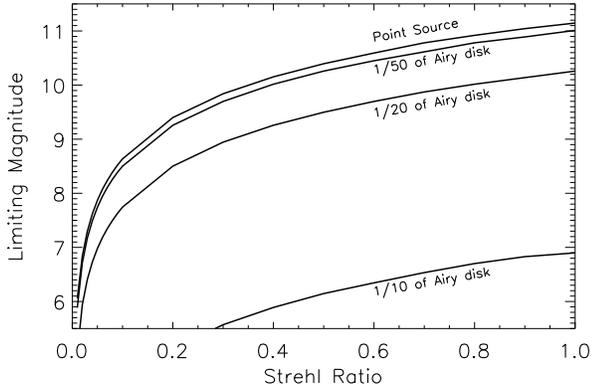}
\caption{\label{fig_mag_2}\footnotesize{Instantaneous limiting
    magnitude of single-mode interferometry technique, defined such that
    the modal SNR per interferogram is  equal to unity. From top to
    bottom we plot limiting magnitudes for object sizes of
    respectively: point source, $1/50$, $1/20$, and $1/10$ of Airy
    disk. Note that Eq. \ref{eq_maglim} reproduces fairly well the curves for
    object sizes equal or smaller than $1/20$ of Airy disk.}}
\end{center}
\end{figure}
\subsection{Performances} \label{sec_AMBER}
In this Section we compute the SNR profiles of the modal visibility (per
interferogram) for faint to bright compact sources (i.e. smaller than
the Airy disk of a single telescope). We consider observations
under average seeing conditions ($r_0=1.6\mathrm{m}$) with 2 Unit
Telescopes ($D=8\mathrm{m}$) at the baselines $Bx=By=100\mathrm{m}$
and the AMBER recombiner in the K band ($2.2\mu m$).
We assume that the interferogram is dispersed
along the columns of a bi-dimensional detector and that each
spectral channel is sampled with 6 pixels to ensure low and high frequency peaks
separation. For quantitative calculations, we choose the specific instrumental 
parameters of AMBER (\cite{malbet_1}, \cite{petrov_etal_1}) together with a spectral
resolution of $35$, an integration time of $30\mathrm{ms}$ per interferogram, a
detector read-out noise of $15\mathrm{e}^{-}/\mathrm{pixel}$ and a
transmission coefficient $\tau = 0.5$. Note that, in those conditions, 
thermal noise is negligible. We also assumed an optimized instrumental
coupling efficiency $\rho_0 = 0.8$ (\cite{shaklan_1}).
Fig. \ref{fig_modal_vis_snr} shows for different Strehl ratios, the modal
visibility SNR as a function of the magnitude for 4 object sizes, 
point source, $1/50$, $1/20$ and $1/10$ of Airy disk, 
with respective visibilities 
$1$, $0.882$,  $0.457$ and  $0.04$ at the baselines
$Bx=By=100\mathrm{m}$.

We can clearly see that the saturation regime, where the modal speckle noise 
dominates, is  significant only in the absence of AO correction. As soon as the
image is partially AO corrected, even at small Strehl ratios
($\mathcal{S} > 0.1$), the saturation regime is rejected towards negative
magnitudes. It is replaced by an extended `photon noise'
regime, which depends on the total flux
weighted by the statistics of the coupling coefficients, that can be interpreted as
transmission coefficients. These transmission coefficients 
decrease with the Strehl ratio and also when the size of the object
increases, therefore lowering the modal visibility SNR. At low fluxes,
the `detector noise' regime, marked by the break in the SNR slope,
takes over.   
\begin{figure*}[!t]
\begin{center}

\includegraphics[width=17cm,height=11cm]{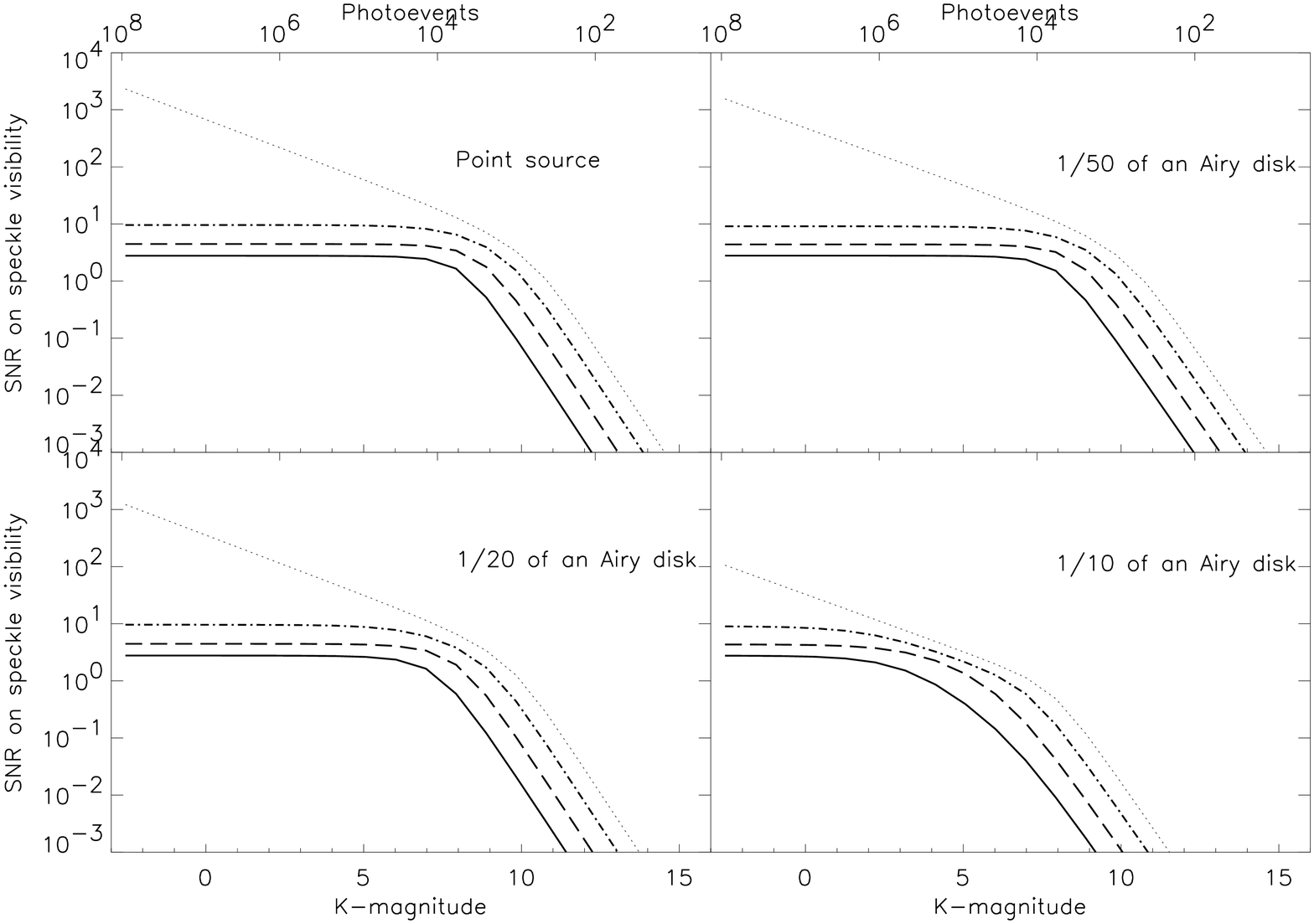}
\caption{\label{fig_speck_vis_snr}\footnotesize{Multispeckle SNR per
    interferogram as a function
    of  the magnitude of the source in the $K$ band. As Fig. \ref{fig_modal_vis_snr}, which those
    curves must be compared to, the object
    size is varying from left to right, top to bottom:
    point source, $1/50$, $1/20$, $1/10$ of Airy disk, for a baseline
    $Bx=By=100\mathrm{m}$.  
    Results are given for different Strehl ratios:
    $\mathcal{S}=0.011$ (solid line, no correction),
    $\mathcal{S}=0.1$ (dashed line), $\mathcal{S}=0.5$ (dash-dotted
    line) and $\mathcal{S}=0.9$ (dotted line).}}
\end{center}
\end{figure*}

We define the instantaneous 
limiting magnitude as the magnitude
for which the SNR per interferogram is equal to unity. It occurs at very
low fluxes, where the additive noise is dominant. From
Eq. \ref{eq_snr_A}, and after some simplifications which are valid for 
partially resolved objects (typically $\sigma_o \ge 1/20$ of Airy disk
at the considered baseline), the corresponding
limiting flux is given by:
%
%
%
%
\begin{equation}
K_{lim}= \frac{\sqrt{2}\sqrt{N_{pix}}\sigma
  N_{tel}}{(1-\tau)V_{ij}\rho_0\mathcal{S}} \label{eq_maglim}
\end{equation}
Figure \ref{fig_mag_2} shows the limiting magnitude as a function of the 
Strehl ratio for four object sizes: point source, $1/50$, $1/20$ and $1/10$. 
Without AO correction, the limiting magnitude is small,
between 5 and 6. As soon as the image is AO corrected and the object
partially resolved, the limiting
magnitude  significantly increases, reaching about $10$ for a Strehl
ratio of $0.5$. However, for largely resolved objects 
($\simeq 1/10$ of Airy disk), 
the coupling efficiency becomes so low that performances of fibered
interferometers in terms of SNR and limiting magnitude are severely degraded.

Note, however, that calculations (their detailed description is
beyond the scope of this paper) show that 
the saturation regime can span higher ranges of
magnitude in cases where the compact
Gaussian source is surrounded by an extended diffuse matter such as a
disk or a dust shell. Indeed, it
can be shown that, depending on the fraction of the flux in the extended
structure, the modal speckle SNR
can decrease  by an order of magnitude, or more. This is
due to the strong fluctuations of the LF coupling coefficient of the extended
component. Such behavior is specific to single-mode interferometry, and
it demonstrates that fibers are more efficient when they deal with
compact sources. On the contrary, we emphasize that in
classical (multimode) interferometry, a diffuse extended component 
has no effect on the SNR. 

\section{Single-Mode versus multimode interferometry} \label{sec_multi}
\begin{figure*}[!t]
\begin{center}
\begin{tabular}{cc}
\includegraphics[width=15cm,height=18cm]{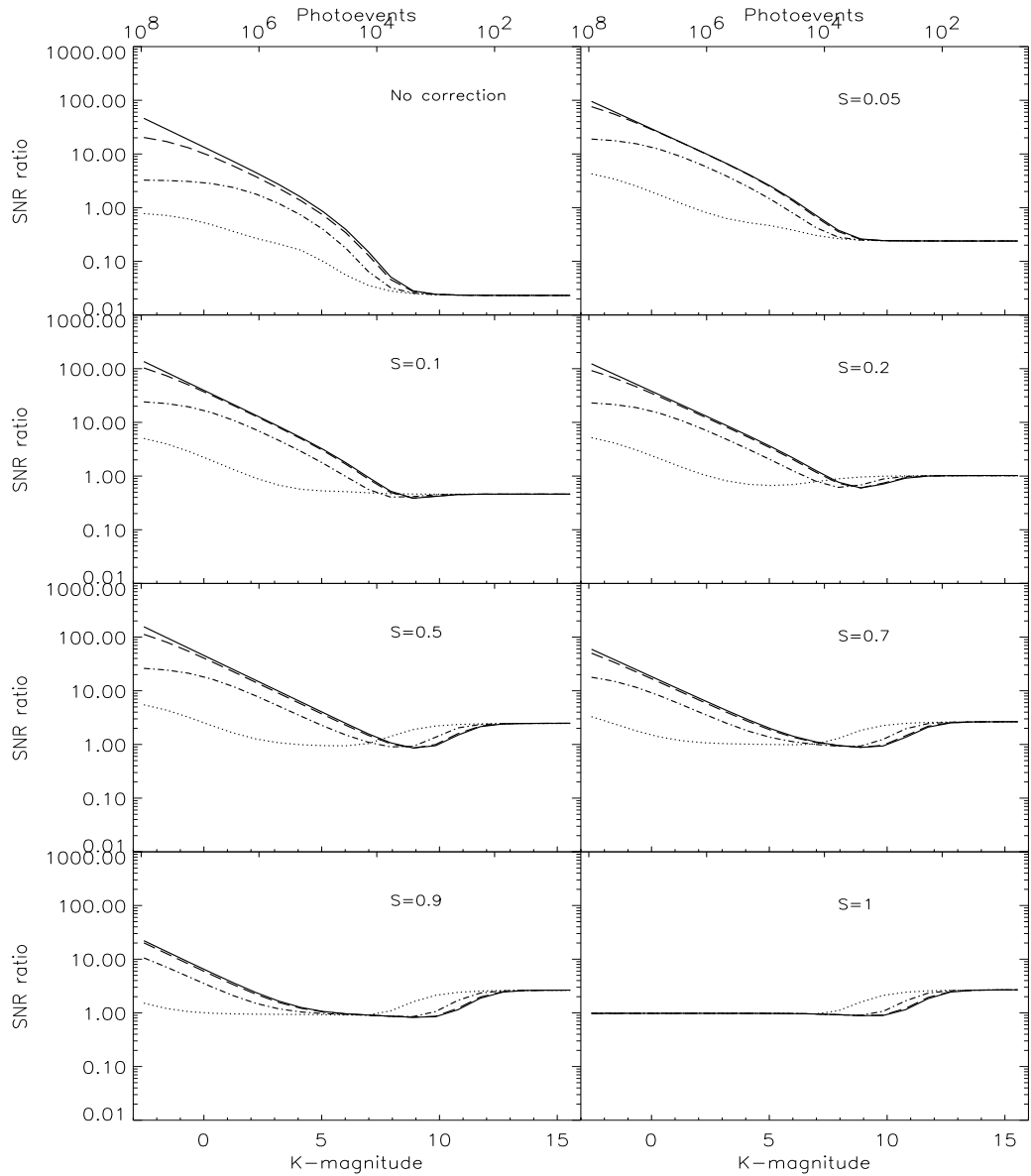}
 \end{tabular}
 \caption{\label{fig_fibers_vs_speckle_snr}\footnotesize{Ratio between
     fibered and multispeckle visibility SNR as a function of the
     magnitude for different AO correction levels, from left to right,
     top to bottom:
    $\mathcal{S} = 0.011$ (no correction),  $\mathcal{S} = 0.05$,  $\mathcal{S} = 0.1$, 
 $\mathcal{S} = 0.2$, $\mathcal{S} = 0.5$,  $\mathcal{S} = 0.7$, $\mathcal{S} = 0.9$ and $\mathcal{S} =1$. Ratios are
     displayed for different source sizes in fraction of Airy disk: 
    point source (solid line), $1/50$ (dashed line),  $1/20$ (dash-dotted
     line),  $1/10$ (dotted line).}}
\end{center}
\end{figure*}

In this Section, we compute the performances of speckle interferometry and
we compare their performances and their robustness to those with single-mode
interferometry.
\subsection{Performances of speckle interferometry}
In absence of waveguides, visibility estimators can be defined by using
speckle techniques, following Labeyrie's method (\cite{labeyrie_1}, \cite{sibille_etal_1})  for
single dish observations (see Appendix D). 
The classical estimator 
(\cite{roddier_lena_1})  consists in taking the ratio
of the integral of the high frequency spectral density by
the integral of the low
frequency one. Taking the integral of the high frequency peak is
essential to perform a consistent comparison with the fiber case, since
fibered interferometry induces an
average (more precisely a convolution) 
of the visibility over the high frequency (see Eq.\ref{eq_rhoij}).
Moreover, to insure a thoroughly consistent comparison, we assume that
the photometric fluxes are measured simultaneously with the
interferograms (instead of taking the integral of the low frequency
spectral density) and we define the estimator of visibility as the ratio
between the integrated high frequency peak of the spectral density and
the photometric fluxes. It may be written as:
\begin{equation}
\widetilde{V^2}(f_{ij}) \propto \frac{\ds
  \int_{f_{ij}-\frac{D}{\lambda}}^{f_{ij}+\frac{D}{\lambda}}|I(f)|^2
  \du{f}}{\ds <k_i k_j>} \label{eq_estim_multi}
\end{equation}
This estimator needs to be calibrated by a point source. It has basically
the same performance than the classical
speckle estimator, but not the same robustness, as we will see later.
In practice, we replace the integral by a discrete sum with a regular
spacing $\Delta f = {r_0 \over \lambda}$. Since we consider
  partial correction by Adaptive Optics, the points involved in the
  discrete sum are not statistically independent. Their correlations are
  taken into account in the  signal to noise ratio calculations,
detailed in Appendix D. These calculations
require the knowledge of the first and second order statistics of the
speckle transfer function. Formal expressions of those moments have been
derived following a  procedure similar to the one described in  Section
4.1. For numerical applications, we adopted the same parameters than for
single-mode interferometry, with the exception of the number of pixels required to correctly sample the interferograms, which is $N=6(D/r_0)$ for 2
telescopes (\cite{chelli_mariotti_1}). 
Fig. \ref{fig_speck_vis_snr} shows, for different Strehl
ratios, the speckle 
visibility SNR as a function of the magnitude, for the four object sizes
previously considered. The striking difference with single-mode
interferometry is the almost absence of the `photon noise' regime, even with
AO correction. Instead, the saturation regime is reached 
in the entire range of magnitudes until the `detector noise' regime takes over. Note, furthermore, that the speckle
noise is barely dependent on the source size and does not cancel out
for a point source, contrarily to the modal speckle noise.
\begin{figure}[!t]
\includegraphics[width=\columnwidth]{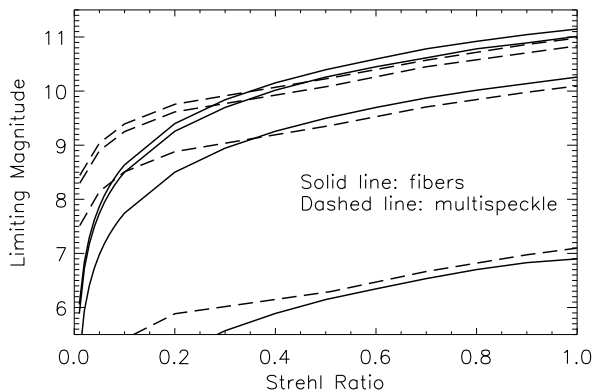}
\caption{\label{fig_mag_3}\footnotesize{Comparison of fibered and
    speckle interferometry limiting magnitudes. Limiting magnitude is
     still defined such that $SNR = 1$. Solid lines and dashed lines
     show fiber and speckle cases respectively. From top to bottom,
     the object size is: point source, $1/50$, $1/20$, and $1/10$ of
     Airy disk.}}
\end{figure}
\subsection{Comparison of performances and robustness} \label{sec_compare}
Fig. \ref{fig_fibers_vs_speckle_snr} shows the ratio between single-mode and speckle SNR as a function of the magnitude, for different compact
source sizes and Strehl ratios. Clearly, with the exception of bright sources
($K < 5$),  speckle interferometry without AO correction is superior to single-mode
interferometry. However, as the degree of AO correction increases, the
situation very rapidly evolves, and, starting from a Strehl ratio of $0.2$,
single-mode interferometry always reaches higher SNR than speckle
interferometry. Such efficiency of fibered interferometry 
is due to two major aspects: the remarkable property of 
spatial filtering of the fiber as soon as enough coherent
energy (typically $20\%$) lies in the Airy disk, together with 
the possibility of sampling the signal on few pixels.
On the contrary, multimode interferometry requires at least good AO 
correction ($\mathcal{S} > 0.5$) to significantly reduce   
the speckle noise.
Moreover, fringe sampling is seeing dependent and requires a much larger
number of pixels, specially for telescopes with large apertures
such as the VLTI. 
Fig. \ref{fig_mag_3} compares the limiting magnitudes of
both methods. Without AO correction, the
speckle limiting magnitude is between 8 and 9, well above the single-mode
one. With AO correction and Strehl ratios larger than 0.2, the limiting
magnitudes are basically the same. 

\begin{figure}[!t]
\includegraphics[width=\columnwidth]{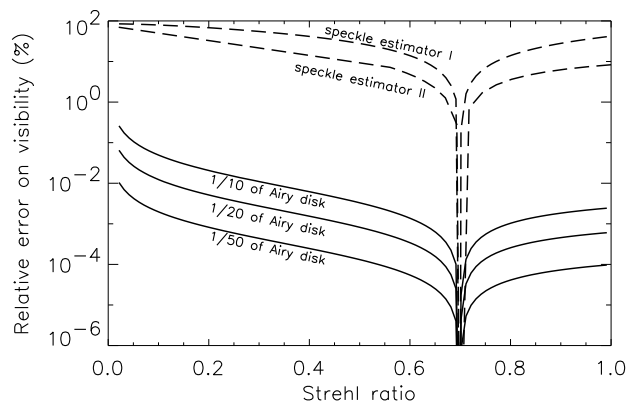}
\caption{\label{fig_robustness}\footnotesize{Robustness of single-mode
    and multimode estimators as a function of the Strehl
    ratio. Comparison is given in terms of relative error on the
    visibility $\Delta{V}/V$. Solid lines deal with fiber estimator
    for respectively $1/10$, $1/20$ and $1/100$ of Airy disk, top to
    bottom. Dashed lines gives the robustness of two speckle
    estimators. Top: so-called estimator I which takes into account
    the estimator defined in this paper (Eq. \ref{eq_estim_multi}). 
    Bottom: estimator II describes the ``classical'' speckle estimator
    (see text). Note that both speckle estimators are object size independent.}}
\end{figure}

To evaluate the robustness of each method, we investigate the stability
of the measured visibility versus the Strehl
ratio. Fig. \ref{fig_robustness} shows the visibility variation 
as a function of
the Strehl ratio, normalized to the visibility at $\mathcal{S} = 0.7$. The
two upper curves correspond to the speckle estimator studied above and to
the classical speckle estimator (ratio of the integrated high and low
frequency peaks of the spectral density), respectively. The classical speckle estimator
is much more robust, but, in both cases, even a small Strehl
ratio variation (0.2) can produce visibility variations up to $10\%$. The
3 lower curves correspond to the modal visibility estimator for source
sizes $1/50$, $1/20$ and $1/10$ of Airy disk (point source is
irrelevant since it is theoretically independent of the
turbulence). 
The robustness of the modal estimator depends on the source size, while the multispeckle estimator does not. In any case, however,
 the modal
estimator is clearly more robust than the speckle one, by more than 2 orders of
magnitude.
The modal visibility is stable at a level less than 1\% over
all the range of possible Strehl ratios, from 0 to 1. This last property is
interesting, not only to perform high precision measurements, but also
for the selection of reference sources. Indeed, it would suggest that the use
of a reference source having a large magnitude difference with the studied object, and hence, where the level of AO correction is different, does not
affect the precision of the measurement.

Note, however, that this comparison focuses on compact sources. For a
central source surrounded by a larger diffuse component, the 
fluctuations of the LF coupling coefficient due to the extended
structure severely reduces the filtering properties of the single-mode
fibers. As a consequence, a situation similar to the multispeckle case
occurs with a saturation
regime spanning a large range of magnitudes, and where the larger the fraction of flux in the extended component, the larger the range. The presence
of an extended structure also causes a decrease of the robustness, which,
however, remains better than in the case of
speckle estimators,  by one order of magnitude.  
\section{Summary}
In this paper, we have developed a formalism that can predict
theoretical SNR on visibilities, when measurements are partially AO
corrected and are corrupted by
photon, additive (detector, thermal) and residual atmospheric noise. This
formalism has been applied to single-mode and multimode (speckle)
interferometers. We have assumed that: (i) the wavefronts over two distinct
  telescopes are uncorrelated; (ii) the
  atmospheric phase has Gaussian statistics, and the associated structure
  function is (spatially) stationary; (iii) the pupil function, the
  object brightness distribution, and the transfer function of the
  turbulent atmosphere are Gaussian. In the case of single-mode interferometry
and for compact sources (i.e. sources smaller than the Airy disk of a
single telescope) not entirely resolved by the interferometer,  we
show that, in
the presence of AO correction, the remarkable filtering properties of 
fibers reject the SNR saturation regime due to speckle noise towards
negative magnitudes. Instead, the modal visibility SNR is dominated by the `photon
noise' regime followed by a break close to the limiting magnitude due to
`detector noise'. For AMBER,
  the near infrared experiment of the VLTI, we found that the limiting magnitude is about $K=10$ for a Strehl ratio of 0.5, with two 8m Unit Telescopes
 (typical atmospheric and instrumental parameters of the
  VLTI/AMBER instrument were discussed in Section
  \ref{sec_AMBER}). In the
case of speckle interferometry, the visibility 
SNR is characterized by the almost absence of the `photon noise' regime,
replaced by the saturation regime due to speckle noise. With the exception
of bright sources ($K < 5$), without AO corrections speckle interferometry reaches
higher SNR than single-mode interferometry, leading to higher limiting
magnitudes. However, the situation is different in presence of AO correction, even small ($\mathcal{S} = 0.2$): the limiting
magnitude of both methods becomes the same, but the single-mode reaches 
much higher SNR than speckle interferometry. 
This last property, together
with the insensitivity of the modal visibility to the Strehl ratio
variations, explains why interferometry with fibers  can produce  visibility
measurement with a very high precision of $1\%$ or less, on
compact sources.

%






\appendix
\section{Moments of the spectral density of an interferogram} \label{app_noise}
\subsection{Formalism}
In order to compute the moments of the spectral density, we use the
spatially continuous model of photodetection process of Goodman (\cite{goodman_1}) where the detected signal takes the form:
\begin{equation}
d(x,y) = \sum_{n=1}^{K}\delta(x-x_n,y-y_n)
\end{equation}
and its Fourier transform: 
\begin{equation}
 \widehat{D}(f_x,f_y) = \sum_{n=1}^{K}\expo^{-2i\pi(f_xx_n+f_yy_n)} \label{eq_dfourier}
\end{equation}
The position ($x_n$, $y_n$) as well as the number of photoevents $K$
are considered as independent random processes with probability
density functions proportional to the intensity $I(x,y)$. This
formalism has been deeply studied (Goodman \cite{goodman_1},
\cite{chelli_1}) and the results already proven will not be
re-demonstrated here. This appendix especially focuses on the original approach where the signal is corrupted by an additive noise $\epsilon$, following a centered Gaussian random process, and of Fourier transform $\widehat{\epsilon}$. Using a mono-dimensional writing, but without loss of generality, the corrupted signal $\widehat{S}(f)$ has now the expression:
\begin{equation}
 \widehat{S}(f) = \widehat{D}(f) + \widehat{\epsilon}(f) = \sum_{n=1}^{K}\expo^{-2i\pi fx_n} + \widehat{\epsilon}(f) \label{eq_noisig}
\end{equation}
We compute the spectral density  statistics with respect
to  the additive Gaussian noise, the photon noise and the turbulent
atmosphere, respectively.
The estimator of the spectral density is:
\begin{equation}
Q(f) = |\widehat{S}(f)|^2
\end{equation}

\subsection{Expected value}
We first take the expected value with respect to the additive noise
from  Eq. \ref{eq_noisig}. It becomes:
\begin{equation}
\meps\{Q(f)\} = |\widehat{D}(f)|^2 + \meps\{|\widehat{\epsilon}|^2\}
= |\widehat{D}(f)|^2 + N\sigma^2
\end{equation} 
Then computing  $|\widehat{D}|^2$ from  Eq. \ref{eq_dfourier}:
\begin{equation}
|\widehat{D}(f)|^2= \sum_{n=1}^{K}\sum_{l=1}^{K}\expo^{-2i\pi(f[x_n-x_l])} \label{eq_unbias_vis}
\end{equation} 
Goodman formalism finally leads to a total expected value of:
\begin{equation}
\mtot\{Q(f)\} = <\overline{K}^2|\widehat{i}(f)|^2>_{\phi} +
<\overline{K}>_{\phi} + N\sigma^2 \label{eq_meanvis}
\end{equation}
where $<>_{\phi}$ denotes the expected value relatively to the turbulent atmosphere random process. 

\subsection{Variance}
The variance is defined by:
\begin{equation}
\sigma^2\{Q\} = \mtot\{Q^2\} - \mtot\{Q\}^2\label{eq_covdef}
\end{equation}
Computing $\mtot\{Q^2\}$ with respect to the additive noise, we have:
\begin{equation}
\mtot\{Q^2\} =  \mtot\{|\widehat{D}|^4\} + 4N\sigma^2\mtot\{|\widehat{D}|^2\} + 2N^2\sigma^4 \label{eq_coefvarvis}
\end{equation}
from which Goodman formalism in the pure photon noise case can be
 applied. Finally the variance of the spectral density may be written as:
\begin{eqnarray}
&&\sigma^2\{Q(f)\} = <\overline{K}^4|\widehat{i}(f)|^4>_{\phi}  - <\overline{K}^2|\widehat{i}(f)|^2>_{\phi} ^2 +\nonumber \\
&& 4<\overline{K}^2|\widehat{i}(f)|^2>_{\phi} - 2<\overline{K}>_{\phi}<\overline{K}^2|\widehat{i}(f)|^2>_{\phi}  +\nonumber \\
&& 4<\overline{K}^2|\widehat{i}(f)|^2>_{\phi} +
2<\overline{K}^2>_{\phi} - <\overline{K}>_{\phi}^2  + 3N\sigma^4 +\nonumber \\
&&N^2\sigma^4 +  2N\sigma^2<\overline{K}>_{\phi} +  2N\sigma^2 <\overline{K}^2|\widehat{i}(f)|^2>_{\phi} 
\end{eqnarray}

\subsection{Covariance}
We can also compute the covariance of the spectral density which is defined as:
\begin{equation}
\mathrm{Cov}\{Q_1,Q_2\} = \mtot\{Q_1Q_2\} - \mtot\{Q_1\}\mtot\{Q_2\}\label{eq_covardef}
\end{equation}
It becomes:
\begin{eqnarray}
&&\mathrm{Cov}\{Q(f_1),Q(f_2)\} =  <\overline{K}^4|\widehat{i}(f_1)|^2|\widehat{i}(f_2)|^2>_{\phi}  - \nonumber\\
&&
<\overline{K}^2|\widehat{i}(f_1)|^2>_{\phi}<\overline{K}^2|\widehat{i}(f_2)|^2>_{\phi}
+ \nonumber\\
&&<\overline{K}^3\widehat{i}(f_1)\widehat{i}^{\ast}(f_2)\widehat{i}^{\ast}(f_1-f_2)>_{\phi}
+\nonumber\\
&&
<\overline{K}^3\widehat{i}^{\ast}(f_1)\widehat{i}(f_2)\widehat{i}(f_1-f_2)>_{\phi}
+  \nonumber \\
&&<\overline{K}^3\widehat{i}(f_1)\widehat{i}(f_2)\widehat{i}^{\ast}(f_1+f_2)>_{\phi}
+\nonumber\\
&&
<\overline{K}^3\widehat{i}^{\ast}(f_1)\widehat{i}^{\ast}(f_2)\widehat{i}(f_1+f_2)>_{\phi}
+  \nonumber \\
&& <\overline{K}^3|\widehat{i}(f_1)|^2>_{\phi} -
<\overline{K}>_{\phi}<\overline{K}^2|\widehat{i}(f_1)|^2>_{\phi}  +
\nonumber\\
&& <\overline{K}^3|\widehat{i}(f_2)|^2>_{\phi} -
<\overline{K}>_{\phi}<\overline{K}^2|\widehat{i}(f_2)|^2>_{\phi}  + \nonumber\\
&& 2<\overline{K}^2|\widehat{i}(f_1)|^2>_{\phi} +
2<\overline{K}^2|\widehat{i}(f_2)|^2>_{\phi} + \nonumber\\
&& <\overline{K}^2|\widehat{i}(f_1-f_2)|^2>_{\phi} +
<\overline{K}^2|\widehat{i}(f_1+f_2)|^2>_{\phi} + \nonumber\\ 
&& <\overline{K}^2>_{\phi}- <\overline{K}>^2_{\phi} + <\overline{K}>_{\phi} + 3N\sigma^4
\end{eqnarray}

\section{Estimator of the visibility for fibered interferometers}\label{app_vis_estimator}
\subsection{The interferometric equation}
M\`ege  (\cite{mege_1}) has shown that the
fibered interferometric equation could be written as follows:  
\begin{eqnarray}
I(f)&=&\sum_i K_i \rho_iH_i(f) +\nonumber \\
&&\sum_i\sum_j \sqrt{K_iK_j}  \rho_{ij}  H_{ij}(f-f_{ij}) \label{eq_interf_ann}
\end{eqnarray}
where $\rho_i$ and $\rho_{ij}$ are respectively the low and high
frequency coupling efficiencies and $K_i$ is the number of detected
photoevents in the absence of a fiber on the $i^{th}$ telescope. Moreover we
define $k_i$ as the number of
photoevents at the output  of the fiber. 
We have the relationship $k_i=\rho_i K_i$. For simplicity
purposes, we also introduce a
``global'' low-frequency coupling coefficient $\rho_{lf}$ such as:
$\rho_{lf} K = \sum_j \rho_j K_j$
where $K$ is the total number of photoevents,
i.e. $K =  \sum_j K_j$.
Supposing that $\forall_i,\overline{\rho}_i=\overline{\rho}$, we have
$\overline{\rho}_{lf} = \overline{\rho}_i = \overline{\rho}$.

\subsection{Estimator of the visibility/Error on the modal visibility}
A classical estimator of the modal
visibility consists in dividing the $HF$ spectral density of the
interferogram by the photometric fluxes. We assume in the following 
that a fraction
$\tau$ of the light has been selected for photometry analysis and that
the rest of the light ($1-\tau$) belongs to the interferogram. 
For sake of simplicity, we
define $K^{\mathcal{P}}$ and $K^{\mathcal{I}}$ as the total number of photoevents
concerning respectively photometric and interferometric channels, i.e
$K^{\mathcal{P}} = \tau K$ and $K^{\mathcal{I}} = (1-\tau) K$. It comes:
\begin{equation}
\widetilde{V_{ij}^2} \propto \frac{<|I^2(f_{ij})|>}{< k^{\mathcal{P}}_i k^{\mathcal{P}}_j>}
\end{equation}
Using Papoulis (\cite{papoulis_1}) second order approximation, we can derive the
square variance relative error of the  modal visibility:
 \begin{eqnarray}
 \frac{\sigma^2\{V_{ij}^2\}}{\overline{V_{ij}^2}^2} &=&
 \frac{\sigma^2\{|I(f_{ij})|^2\}}{\mtot^2\{|I(f_{ij})|^2\}}
 + 
 \frac{\sigma^2\{k^{\mathcal{P}}_i k^{\mathcal{P}}_j\}}{\overline{k^{\mathcal{P}}_i}^2\overline{k^{\mathcal{P}}_j}^2}
 \nonumber \\
 &&-2\frac{\mathrm{Cov}\{|I(f_{ij})|^2,k^{\mathcal{P}}_ik^{\mathcal{P}}_j\}}{\mtot\{|I(f_{ij})|^2\}\overline{k^{\mathcal{P}}_i}\overline{k^{\mathcal{P}}_j}}
 \end{eqnarray}
To derive those moments, we use  the Goodman formalism described in
Appendix \ref{app_noise}. In the case of fibered interferometers, the
$LF$ and $HF$ of the Fourier-transformed interferogram ($H_i(f)$ and
$H_{ij}(f)$) are fixed by
the geometry of the fibers, and statistics
with respect to the turbulent atmosphere only appear in the coupling
coefficients. For sake of simplicity we assume those coupling
coefficients to be uncorrelated between two different baselines, although
it does not change the eventual conclusions. Assuming also that the telescope
transmissions are all equal, i.e. $K_i = K/N_{tel}$,  
it leads to the following expressions:
\begin{eqnarray}
&&\mtot\{|I(f_{ij})|^2\} =
\overline{|\rho_{ij}|^2}\frac{\overline{K^{\mathcal{I}}}^2}{N_{tel}^2}+
\overline{\rho}\overline{K^{\mathcal{I}}} +  N\sigma^2 
\end{eqnarray}
\begin{eqnarray}
\sigma^2\{|I(f_{ij})|^2\} &=&
\sigma^2_{\rho_{ij}^2}\frac{\overline{K^{\mathcal{I}}}^4}{N_{tel}^4}+\left[2\overline{\rho_{lf}^2}-\overline{\rho}^2\right]\overline{K^{\mathcal{I}}}^2+  \nonumber \\
&& \left[4\overline{\rho_{lf}|\rho_{ij}|^2}-2\overline{\rho}\overline{|\rho_{ij}|^2}\right]\frac{\overline{K^{\mathcal{I}}}^3}{N_{tel}^2} 
  + \nonumber \\
&& 4\overline{|\rho_{ij}|^2}\frac{\overline{K^{\mathcal{I}}}^2}{N_{tel}^2} + 2N\sigma^2
 \overline{|\rho_{ij}|^2}\frac{\overline{K^{\mathcal{I}}}^2}{N_{tel}^2} + \nonumber \\
&&  2N\sigma^2\overline{\rho}\overline{K^{\mathcal{I}}} +  3N\sigma^4 + N^2\sigma^4 
\end{eqnarray}
\begin{eqnarray}
\sigma^2\{k^{\mathcal{P}}_i k^{\mathcal{P}}_j\} &=&
\sigma^2_{\rho_i\rho_j} \frac{ \overline{K^{\mathcal{P}}}^4}{N_{tel}^4}+
2\overline{\rho^2}\overline{\rho}  \frac{\overline{K^{\mathcal{P}}}^3}{N_{tel}^3}
 +  \overline{\rho}^2\frac{\overline{K^{\mathcal{P}}}^2}{N_{tel}^2}
\end{eqnarray}
{\small \begin{eqnarray}
 \mathrm{Cov}\{|I(f_{ij})|^2,k^{\mathcal{P}}_i k^{\mathcal{P}}_j\} = \frac{\overline{K^{\mathcal{P}}}^2\overline{K^{\mathcal{I}}}^2}{N_{tel}^4}\left[\overline{|\rho_{ij}|^2\rho_i\rho_j}-\overline{|\rho_{ij}^2|}\overline{\rho}^2\right]
\end{eqnarray}}
And the square relative error on the modal visibility can be expressed as the
sum of three contributions: 
\begin{equation}
\frac{\sigma^2\{V_{ij}^2\}}{\overline{V_{ij}^2}^2} = \mathcal{E}^2_P(K,\rho) 
+ \mathcal{E}^2_A(K, \sigma^2_{\mathcal{A}}, \rho) + \mathcal{E}^2_S(\rho)
\end{equation}
where:
\begin{eqnarray}
\mathcal{E}^2_{P}
&=&\left[\frac{N_{tel}(4\overline{\rho_{lf}|\rho_{ij}|^2}-2\overline{\rho}\overline{|\rho_{ij}|^2})}{(1-\tau)\overline{|\rho_{ij}|^2}^2}
  +\frac{2}{\tau}\frac{\overline{\rho^2}}{\overline{\rho}^3}\right]\frac{N_{tel}}{\overline{K}}
\nonumber \\
& + &
\left[\frac{N_{tel}^2(2\overline{\rho_{lf}^2}-\overline{\rho}^2)}{(1-\tau)^2\overline{|\rho_{ij}|^2}^2}
  +\frac{4}{(1-\tau)^2\overline{|\rho_{ij}|^2}}
  +\frac{1}{\tau^2\overline{\rho}^2}\right]
\frac{N_{tel}^2}{{\overline{K}^2}}\nonumber \\
& + & \frac{\overline{\rho}}{(1-\tau)^3\overline{|\rho_{ij}|^2}^2} \frac{N_{tel}^4}{{\overline{K}^3}}
\end{eqnarray}
is the photon noise square relative error:
\begin{eqnarray}
\mathcal{E}^2_A &=&
\frac{3N\sigma^4 + N^2\sigma^4}{(1-\tau)^4\overline{|\rho_{ij}|^2}^2}\frac{N_{tel}^4}{\overline{K}^4}
+
\frac{2N\sigma^2}{(1-\tau)^2\overline{|\rho_{ij}|^2}}\frac{N_{tel}^2}{\overline{K}^2}
\nonumber \\
&+& \frac{2N\sigma^2\overline{\rho}}{(1-\tau)^3\overline{|\rho_{ij}|^2}^2}\frac{N_{tel}^4}{\overline{K}^3}
\end{eqnarray}
takes into account the additive noise $\sigma^2$ and
\begin{eqnarray}
&&\mathcal{E}^2_S =
\frac{\sigma^2_{\rho_{ij}^2}}{\overline{|\rho_{ij}|^2}^2}+\frac{\sigma^2_{\rho_i\rho_j}}{\overline{\rho}^4}-2\frac{\mathrm{Cov}\{\rho_{ij}^2,\rho_i\rho_j\}}{\overline{|\rho_{ij}|^2}\overline{\rho}^2}
\end{eqnarray}
arises from the coupling fluctuations inducing a so called
``modal-speckle'' square relative error. 
Concerning the latter contribution, it is interesting to notice that in the
case of a point source, we have $|\rho_{ij}|^2=\rho_{i}\rho_{j}$.
The outcome is that the ``modal-speckle'' noise contribution for a point
source is zero.

\section{Coupling coefficients statistics} \label{app_pseudo_speck}
\begin{table*}[!t]
\caption{\label{table_exp_coupling_before}\footnotesize{LF and HF coupling
    coefficient  first and second order statistics. Rigorous expressions.}}
\begin{eqnarray}
\hline \nonumber \\
\overline{\rho_{i}} & = & \frac{\rho_0}{S}\int
V_{\star}^{\ast}(u)P(\alpha)P(\alpha+u)<\Psi_i(\alpha)\Psi_i^{\ast}(\alpha+u)>
\du\alpha\du{u} \label{eq_app_fib1_before}\\  
\overline{|\rho_{ij}|^2} &  = & \frac{\rho_0^2}{S^2}\int
V_{\star}^{\ast}(u-f_{ij})V_{\star}(v-f_{ij})P(\alpha)P(\alpha+u)P(\beta)P(\beta+v)<\Psi_i(\alpha)\Psi_i^{\ast}(\beta)\Psi_j^{\ast}(\alpha+u)\Psi_j(\beta+v)>
\du\alpha\du{u} \du\beta\du{v} \label{eq_app_fib2_before}\\
\hline \nonumber 
\end{eqnarray}
\begin{eqnarray}
\hline \nonumber \\
&&\mathrm{E}\{|\rho_{ij}|^4\} = \nonumber\\ 
&&\frac{\rho_0^4}{S^4}\int
V_{\star}(f_{ij}-a_1)V_{\star}^{\ast}(f_{ij}-a_2)V_{\star}(f_{ij}-a_3)V_{\star}^{\ast}(f_{ij}-a_4)P(\alpha)P(\alpha+a_1)P(\beta)P(\beta+a_2)P(\gamma)P(\gamma+a_3)P(\delta)P(\delta+a_4) \nonumber \\
&&<\Psi_i(\alpha)\Psi_i^{\ast}(\beta)\Psi_i(\gamma)\Psi_i^{\ast}(\delta)\Psi_j^{\ast}(\alpha+a_1)\Psi_j(\beta+a_2)\Psi_j^{\ast}(\gamma+a_3)\Psi_j(\delta+a_4)>
\du{a_1}\du{a_2}\du{a_3}\du{a_4}\du{\alpha}\du{\beta}\du{\gamma}\du{\delta}  \label{eq_app_fib3_before} \\ 
&&\mathrm{E}\{\rho_i^2\rho_j^2\} = \nonumber \\
&&\frac{\rho_0^4}{S^4}\int V_{\star}^{\ast}(a_1)V_{\star}^{\ast}(a_2)V_{\star}^{\ast}(a_3)V_{\star}^{\ast}(a_4)P(\alpha)P(\alpha+a_1)P(\beta)P(\beta+a_2)P(\gamma)P(\gamma+a_3)P(\delta)P(\delta+a_4)\nonumber\nonumber\\
&&<\Psi_i(\alpha)\Psi_i^{\ast}(\alpha+a_1)\Psi_i(\beta)^{\ast}\Psi_i(\beta+a_2)\Psi_j(\gamma)\Psi_j^{\ast}(\gamma+a_3)\Psi_j(\delta)\Psi_j^{\ast}(\delta+a_4)>
\du{a_1}\du{a_2}\du{a_3}\du{a_4}\du{\alpha}\du{\beta}\du{\gamma}\du{\delta}
\label{eq_app_fib4_before}\\ 
&&\mathrm{E}\{|\rho_{ij}|^2\rho_i\rho_j\} = \nonumber\\
&&\frac{\rho_0^4}{S^4}\int
V_{\star}(f_{ij}-a_1)V_{\star}^{\ast}(f_{ij}-a_2)V_{\star}^{\ast}(a_3)V_{\star}^{\ast}(a_4) P(\alpha)P(\alpha+a_1)P(\beta)P(\beta+a_2)P(\gamma)P(\gamma+a_3)P(\delta)P(\delta+a_4)\nonumber \\
&& <\Psi_i(\alpha)\Psi_i^{\ast}(\beta)\Psi_i(\gamma)\Psi_i^{\ast}(\gamma+a_3)\Psi_j^{\ast}(\alpha+a_1)\Psi_j(\beta+a_2)\Psi_j(\delta)\Psi_j^{\ast}(\delta+a_4)>
\du{a_1}\du{a_2}\du{a_3}\du{a_4}\du{\alpha}\du{\beta}\du{\gamma}\du{\delta}\label{eq_app_fib5_before} \\
\hline \nonumber 
\end{eqnarray}
\end{table*}

In order to calculate the square relative error of the modal
visibility given in Appendix \ref{app_vis_estimator}, we first compute
first and second order
moments of the $LF$ and $HF$ coupling coefficients which are defined
as following:
\begin{eqnarray}
\rho_i(V_{\star})&=&\rho_0 \frac{(V_{\star}*T^i)_{f=0}}{\int T^{i}_0(f) \du{f}}  \label{eq_rhoi_ann} \\ 
\rho_{ij}(V_{\star})&=&\rho_0
\frac{(V_{\star}*T^{ij})_{f=f_{ij}}}{\sqrt{\int T^{i}_0(f) \du{f}\int
    T^{j}_0(f) \du{f}}}  \label{eq_rhoij_ann}
\end{eqnarray}
where $T^{i}$ and $T^{ij}$ are the (partially AO corrected) modal
transfer function resulting respectively from the normalized
auto-correlation and cross-correlation of the entrance 
aberration-corrupted pupil weighted by the fiber single mode, i.e:
\begin{eqnarray}
T^{i}(u) &=& \frac{1}{S}\int P_i(\alpha)P_i(\alpha+u)\Psi_i(\alpha)\Psi_i^{\ast}(\alpha+u) \du\alpha \label{eq_transf1_ann}\\
T^{ij}(u) &=&  \frac{1}{S}\int P_i(\alpha)P_j(\alpha+u)\Psi_i(\alpha)\Psi_j^{\ast}(\alpha+u) \du\alpha \label{eq_transf2_ann}
\end{eqnarray}
Introducing  Eq.'s \ref{eq_transf1_ann}, \ref{eq_transf2_ann}
in Eq.'s \ref{eq_rhoi_ann} and \ref{eq_rhoij_ann} and developing the
expressions of first order ($\overline{\rho_{i}}$,
$\overline{|\rho_{ij}|^2}$), and second order 
($\sigma^2_{|\rho_{ij}|^2}$, $\sigma^2_{\rho_i\rho_j}$,
$\mathrm{Cov}\{|\rho_{ij}|^2,\rho_i\rho_j\}$) coupling coefficient statistics, 
leads respectively to
second, fourth and  eighth order moments of the complex amplitude of
the wavefronts $\Psi_i(u)$ and $\Psi_j(u)$. Such rigorous expressions are written in Table
\ref{table_exp_coupling_before}.

At this point, those expressions are not yet formally computable. Hence
we  perform simplifications of the equations. We first assume that the wavefronts
are uncorrelated,
i.e.
$<\Psi_i(u)\Psi_j(u)>_{\phi}=<\Psi_i(u)>_{\phi}<\Psi_j(u)>_{\phi}$. Then
we use Korff's (\cite{korff_1}) derivation of the moments of the complex
amplitude of the wavefronts to introduce in the
equations, linear combinations of the structure function
($\mathcal{D}(u,v)$) at different
spatial frequencies. We recall that the structure function is defined
such that (\cite{conan_1}):
\begin{equation}
\mathcal{D}(u,v) = <\psi_i(u)\psi_i^{\ast}(v)>_{\phi} = <\psi_j(u)\psi_j^{\ast}(v)>_{\phi}
\end{equation}
Finally we suppose that
these structure functions are stationary, i.e. $\mathcal{D}(u,v) =
\mathcal{D}(u-v)$. After such formal derivations we obtain expressions of the coupling
coefficients statistics as summarized in Tables \ref{table_exp_coupling}.
Readers may note that in the pure turbulent case, we assume in
addition  that the complex amplitude of
the wavefront follows circular Gaussian statistics
(\cite{roddier_1}), hence slightly changing the expressions given in
table \ref{table_exp_coupling}
\begin{table*}[!t]
\caption{\label{table_exp_coupling}\footnotesize{LF and HF coupling
    coefficient  first and second order statistics. simplified
    expressions after assuming: (i) no correlation between wavefronts; 
    (ii) Korff's derivation of the moments of the complex amplitude of
    the wavefronts; (iii) stationarity of the
    structure function}}
\begin{eqnarray}
\hline \nonumber \\
\overline{\rho_{i}} & = & \frac{\rho_0}{S}\int
V_{\star}^{\ast}(u)P(\alpha)P(\alpha+u)\mathrm{e}^{-\frac{1}{2}\mathcal{D}(u)}
\du\alpha\du{u} \label{eq_app_fib1}\\  
\overline{|\rho_{ij}|^2} &  = & \frac{\rho_0^2}{S^2}\int
V_{\star}^{\ast}(u-f_{ij})V_{\star}(v-f_{ij})P(\alpha)P(\alpha+u)P(\beta)P(\beta+v)
\mathrm{e}^{-\frac{1}{2}[\mathcal{D}(\beta-\alpha)-\mathcal{D}(\beta-\alpha+v-u)]}
\du\alpha\du{u} \du\beta\du{v}   \label{eq_app_fib2}\\ 
\hline \nonumber \\ \nonumber 
\end{eqnarray}
\begin{eqnarray}
\hline \nonumber \\
&&\mathrm{E}\{|\rho_{ij}|^4\} = \nonumber\\ 
&&\frac{\rho_0^4}{S^4}\int
V_{\star}(f_{ij}+a_1-a_2)V_{\star}^{\ast}(f_{ij}+a_3-a_4)V_{\star}(f_{ij}+a_5-a_6)V_{\star}^{\ast}(f_{ij}+a_7-a_8)P(a_1)P(a_2)P(a_3)P(a_4)
P(a_5)P(a_6)P(a_7)P(a_8) \nonumber \\
&&\mathrm{e}^{-\frac{1}{2}[\mathcal{D}(a_1-a_3)+\mathcal{D}(a_5-a_7)+\mathcal{D}(a_2-a_4)+\mathcal{D}(a_6-a_8)]}\left[\frac{\mathrm{e}^{-\frac{1}{2}[\mathcal{D}(a_3-a_5)+\mathcal{D}(a_4-a_6)+\mathcal{D}(a_1-a_7)+\mathcal{D}(a_2-a_8)]}}{\mathrm{e}^{-\frac{1}{2}[\mathcal{D}(a_1-a_5)+\mathcal{D}(a_2-a_6)+\mathcal{D}(a_3-a_7)+\mathcal{D}(a_4-a_8)]}}\right]
\du{a_1}\ldots\du{a_8} \label{eq_app_fib3} \\ 
&& \mathrm{E}\{\rho_i^2\rho_j^2\} = \nonumber\\
&&\frac{\rho_0^4}{S^4}\int V_{\star}^{\ast}(a_2-a_1)V_{\star}^{\ast}(a_4-a_3)V_{\star}^{\ast}(a_6-a_5)V_{\star}^{\ast}(a_8-a_7)P(a_1)P(a_2)P(a_3)P(a_4) P(a_5)P(a_6)P(a_7)P(a_8)\nonumber\\
&&\mathrm{e}^{-\frac{1}{2}[\mathcal{D}(a_2-a_1)+\mathcal{D}(a_4-a_3)+\mathcal{D}(a_6-a_5)+\mathcal{D}(a_8-a_7)]}\left[\frac{\mathrm{e}^{-\frac{1}{2}[\mathcal{D}(a_2-a_3)+\mathcal{D}(a_1-a_4)+\mathcal{D}(a_6-a_7)+\mathcal{D}(a_5-a_8)]}}{\mathrm{e}^{-\frac{1}{2}[\mathcal{D}(a_3-a_1)+\mathcal{D}(a_4-a_2)+\mathcal{D}(a_5-a_7)+\mathcal{D}(a_8-a_6)]}}\right]
\du{a_1}\ldots\du{a_8} \label{eq_app_fib4}\\ 
&&\mathrm{E}\{|\rho_{ij}|^2\rho_i\rho_j\} = \nonumber\\
&&\frac{\rho_0^4}{S^4}\int
V_{\star}(f_{ij}+a_1-a_2)V_{\star}^{\ast}(f_{ij}+a_3-a_4)V_{\star}^{\ast}(a_6-a_5)V_{\star}^{\ast}(a_8-a_7)P(a_1)P(a_2)P(a_3)P(a_4)
P(a_5)P(a_6)P(a_7)P(a_8) \nonumber \\
&&\mathrm{e}^{-\frac{1}{2}[\mathcal{D}(a_3-a_1)+\mathcal{D}(a_2-a_4)+\mathcal{D}(a_6-a_5)+\mathcal{D}(a_8-a_7)]}\left[\frac{\mathrm{e}^{-\frac{1}{2}[\mathcal{D}(a_3-a_5)+\mathcal{D}(a_6-a_1)+\mathcal{D}(a_7-a_2)+\mathcal{D}(a_8-a_4)]}}{\mathrm{e}^{-\frac{1}{2}[\mathcal{D}(a_5-a_1)+\mathcal{D}(a_6-a_3)+\mathcal{D}(a_7-a_4)+\mathcal{D}(a_8-a_2)]}}\right]
\du{a_1}\ldots\du{a_8} \label{eq_app_fib5} \\
\hline \nonumber 
\end{eqnarray}
\end{table*}

\section{Estimator of the visibility for multispeckle interferometry} 
\label{app_multispeck}
\subsection{The interferometric equation}
For non fibered interferometers, the convolution between the object
and the interferometer transfer function stands for any spatial
frequency (\cite{tallon_1}). We may express: 
\begin{equation}
I(f) = \sum_i K_i i(f) +  \sum_i\sum_j \sqrt{K_iK_j} i(f-f_{ij})
\end{equation}
with 
\begin{equation}
i(f) = V(f).S(f)
\end{equation}
where $V(f)$ is the object visibility  and $S(f)$ is the normalized transfer
function of the interferometer which consists in the autocorrelation
of the pupil function weighted by the remaining atmospheric phase.
We have the following relationship 
\begin{equation}
S(f) = \sum_i S^{i}(f) + \frac{1}{N_{tel}} \sum_i\sum_j S^{ij}(f)
\end{equation}
where $S^{ii}(f)$ and $S^{ij}(f)$ are respectively the auto and
cross-correlation of single phase corrugated pupil (\cite{roddier_1}), i.e.:
\begin{eqnarray}
S^{ii}(f) &=& \frac{1}{S} \int  P(\alpha)P(\alpha+f)\Psi_i(\alpha)\Psi_i^{\ast}(\alpha+f) \du\alpha \\
S^{ij}(f) &=& \frac{1}{S} \int  P(\alpha)P(\alpha+f)\Psi_i(\alpha)\Psi_j^{\ast}(\alpha+f) \du\alpha 
\end{eqnarray}
\subsection{Estimator of the visibility/Error on the visibility}
To perform a consistent comparison within the fiber case, we define
the estimator of the multispeckle visibility as the ratio between the
integrated high frequency peak and the photometric fluxes. We still
assume that a fraction $\tau$ of the light is injected in the
photometric channels. It writes:  
\begin{equation}
\widetilde{V^2}(f_{ij}) \propto \frac{\ds
  \int_{f_{ij}-\frac{D}{\lambda}}^{f_{ij}+\frac{D}{\lambda}}|I(f)|^2
  \du{f}}{\ds <k^{\mathcal{P}}_ik^{\mathcal{P}}_j>}
\end{equation}
From a digital point of view, as the size of the instantaneous
speckle image is $\frac{\lambda}{r_0}$ wide (perfect correction
apart), the integral of the previous equation can be replaced 
by a discrete sum over points of
the peak that are distributed every $\frac{r_0}{\lambda}$. We then have
an equivalent estimator:
\begin{equation}
\widetilde{V^2}(f_{ij}) \propto \frac{\ds
  \sum_{l=f_{ij}-\frac{D}{\lambda}}^{f_{ij}+\frac{D}{\lambda}}|I(l)|^2}{\ds
  <k^{\mathcal{P}}_ik^{\mathcal{P}}j>}, \Delta{l}=\frac{r_0}{\lambda} 
\end{equation}
We derive the square relative error from Papoulis (\cite{papoulis_1}):
\begin{eqnarray}
\frac{\sigma^2\{V^2_{ij}\}}{\overline{V^2_{ij}}^2} &=& \frac{\ds
  \sum_{l}\sigma^2\{|I(l)|^2\}
  + \sum_{m,n \neq m}\mathrm{Cov}\{|I(m)|^2|I(n)|^2\}} {\left[\ds
  \sum_{l}\mtot\{|I(l)|^2\}\right]^2}  \nonumber \\
&& + \frac{1}{k^{\mathcal{P}}_i} + \frac{1}{k^{\mathcal{P}}_j} + \frac{1}{k^{\mathcal{P}}_ik^{\mathcal{P}}_j} \label{eq_error_multispeckle}
\end{eqnarray}
Note that correlations between points have to be
  taken into account since we consider partial correction by
  Adaptive Optics.
We then use the formalism developed in Appendix \ref{app_noise} to
derive \ref{eq_error_multispeckle}.
Note that in the
multispeckle case, at the contrary of the fibered case, 
 the $LF$ and $HF$ peaks of the
 Fourier-transformed interferogram (i.e. $ S^{i}(f)$ and $S^{ij}
(f)$)
 depend on the turbulence.  Moments of $|I(l)|^2$ are:
\begin{eqnarray}
&&\mtot\{|I(l)|^2\} = 
\frac{\overline{K^{\mathcal{I}}}^2}{N_{tel}^2}<|S^{ij}(l)|^2>V^2_{l} + \overline{K^{\mathcal{I}}}
+ N\sigma^2 
\end{eqnarray}
\begin{eqnarray}
 &&\sigma^2\{|I(l)|^2\} =  \overline{K^{\mathcal{I}}} + \overline{K^{\mathcal{I}}}^2 + \nonumber \\
&& 
\left[2\frac{\overline{K^{\mathcal{I}}}^3}{N_{tel}^2}+4\frac{\overline{K^{\mathcal{I}}}^2}{N_{tel}^2}\right]<|S^{ij}(l)|^2>_{\phi}V^2_{l}+ \nonumber \\
&&\frac{\overline{K^{\mathcal{I}}}^4}{N_{tel}^4}\bigg[<|S^{ij}(l)|^4>_{\phi}-<|S^{ij}(l)|^2>_{\phi}^2\bigg]V^4_{l} +
\nonumber
\\
&&2N\sigma^2\frac{\overline{K^{\mathcal{I}}}^2}{N_{tel}^2}<|S^{ij}(l)|^2>_{\phi}V^2_{l}
+ 2N\sigma^2\overline{K^{\mathcal{I}}} +\nonumber \\
&& 3N\sigma^4 + N^2\sigma^4
\end{eqnarray}
\begin{eqnarray}
&&\mathrm{Cov}\{|I(m)|^2|I(n)|^2\} = \nonumber\\
&&\overline{K^{\mathcal{I}}} +\frac{\overline{K^{\mathcal{I}}}^2}{N_{tel}^2}\left[2<|S^{ij}(m)|^2>_{\phi}V^2_{m}  + \right.
 \nonumber \\
&& \left. 
  2<|S^{ij}(n)|^2>_{\phi}V^2_{n} +  <|S^{i}(m-n)|^2>_{\phi}V^2_{m-n}\right] +  \nonumber \\
&&2\frac{\overline{K^{\mathcal{I}}}^3}{N_{tel}^2}<S^{ij}(m)S^{ij}(n)S^{i^{\ast}}(m-n)> V_{m}V_{n}V^{\ast}_{m-n}+
\nonumber \\
&&\frac{\overline{K^{\mathcal{I}}}^4}{N_{tel}^4}\bigg[<|S^{ij}(m)|^2|S^{ij}(n)|^2>_{\phi}- \nonumber \\
&&<|S^{ij}(m)|^2>_{\phi}<|S^{ij}(n)|^2>_{\phi}\bigg]V^2_{m}V^2_{n}
+ 3N\sigma^4
\end{eqnarray}

Another classical estimator of the visibility (\cite{roddier_lena_1},
\cite{mourard_etal_1}), known to be more robust to the turbulence, 
consists in taking the ratio 
of the high-frequency peak energy by the low-frequency one (coherent
energy versus incoherent one):
\begin{equation}
\widetilde{V^2}(f_{ij}) = \frac{\ds \int_{f_{ij}-\frac{D}{\lambda}}^{f_{ij}+\frac{D}{\lambda}}|I(f)|^2 \du{f}}{\ds \int_{-\frac{D}{\lambda}}^{\frac{D}{\lambda}}|I(f)|^2 \du{f}}
\end{equation}
Assuming that photometric channels
contain at most $50\%$ of the information, we
expect the relative error of the visibility from this estimator to be
equal to the previous one within a factor of $\sqrt{2}$.

\section{Speckle transfer function statistics}
\label{app_multispeck_stats}
\begin{table*}[*!t]
\caption{\label{table_esp_transfunc}\footnotesize{Formal expressions
    of respectively LF and HF speckle transfer function first and
    second order statistics.}}
 \begin{eqnarray}
 \hline \nonumber  \\ 
<|S^{i}(f)|^2> &=&  S^{-2} \int P(a_1)P(a_1+a_2)P(a_1+f)P(a_1+a_2+f)
\left[\frac{\mathrm{e}^{-[\mathcal{D}(a_2)+\mathcal{D}(f)]}}{\mathrm{e}^{-\frac{1}{2}[\mathcal{D}(a_2+f)+\mathcal{D}(a_2-f)]}}\right]
\du{a_1}\du{a_2} \label{eq_app_speck1}\\
<|S^{ij}(f)|^2> &=&  S^{-2} \int P(a_1)P(a_2)P(a_1+f)P(a_2+f)
\mathrm{e}^{-\mathcal{D}(a_1-a_2)} \du{a_1}\du{a_2}
\label{eq_app_speck2} \\
\hline \nonumber 
\end{eqnarray}
 \begin{eqnarray}
\hline \nonumber  \\ 
&&<|S^{ij}(f)|^4> = \nonumber \\
&&S^{-4} \int P(a_1)P(a_1+f)P(a_2)P(a_2+f)P(a_1+a_3)P(a_1+a_3+f)P(a_2+a_4)P(a_2+a_4+f) \nonumber \\
&&\left[\frac{\mathrm{e}^{-[\mathcal{D}(a_1-a_2)+\mathcal{D}(a_3+a_1-a_2)+\mathcal{D}(a_2+a_4-a_1)+\mathcal{D}(a_4+a_2-a_1-a_3)]}}{\mathrm{e}^{-[\mathcal{D}(a_3)+\mathcal{D}(a_4)]}}\right]
\du{a_1}\du{a_2}\du{a_3}\du{a_4} \label{eq_app_speck4} \\  \hline \nonumber 
\end{eqnarray}
\end{table*}

In order to calculate the square relative error of the visibility, we
first need to compute expected values of $|S^{i}(f)|^2$,
$|S^{ij}(f)|^2$ and $|S^{ij}(f)|^4$. As discussed in Appendix
\ref{app_multispeck} the functions $S^{i}(f)$ and $S^{i}(f)$ are respectively the auto and the 
cross-correlation
of the pupil functions weighted by the remaining atmospheric
wavefront. If we neglect the weighting of the pupil by the single mode
of the fiber, we find the very same expression as in Eq's
\ref{eq_transf1_ann}, \ref{eq_transf2_ann}, i.e. $S^{i}(u) =
T^{i}(u)$ and $S^{ij}(u) = T^{ij}(u)$.
Deriving $<|S^{i}(f)|^2>$,
$<|S^{ij}(f)|^2>$,  $<|S^{i}(f)|^4>$ , $<|S^{ij}(f)|^4>$ introduces
once again second, fourth and  eighth order statistics of the complex 
amplitude of the wavefronts, that we develop following the same
procedure as described in Appendix \ref{app_pseudo_speck}. 
We finally obtained expressions given in Table
\ref{table_esp_transfunc}.



\end{document}